\documentstyle[12pt]{article}
\normalbaselines
\begin{document}
\thispagestyle{empty}
\noindent \begin{center}\Large Charged Particle with Magnetic Moment
in the Aharonov-Bohm Potential\end{center}
\vspace{1cm}
\centerline{M. Bordag and S. Voropaev\footnote{Permanent address: Vernadsky
Institute, Lab. of Theoretical and Mathematical Physics, Kos\-sygin street 19,
Moscow, Russia}}
\vspace{1cm}
\centerline{Universit\"at Leipzig, FB Physik}
\centerline{Augustusplatz 10}
\centerline{O-7010 Leipzig, Germany}
\vspace{1cm}

\section*{Abstract}

We considered a charged quantum mechanical particle with spin
${1\over 2}$ and gyromagnetic ratio $g\ne 2$ in the field af a
magnetic string. Whereas the interaction of the charge with the
string is the well kown Aharonov-Bohm effect and the
contribution of magnetic moment associated with the spin in the
case $g=2$ is known to yield an additional scattering and zero
modes (one for each flux quantum), an anomaly of the magnetic
moment (i.e. $g>2$) leads to bound states. We considered two
methods for treating the case $g>2$.

The first is the method of self adjoint extension of the corresponding
Hamilton operator. It yields one bound state as well as additional scattering.
In the second we consider three exactly solvable models for finite flux
tubes and take the limit of shrinking its radius to zero. For finite radius,
there are $N+1$ bound states ($N$ is the number of flux quanta in the tube).

For $R\to 0$ the bound state energies tend to infinity so that this limit is
not physical
unless $g\to 2$ along with $R\to 0$. Thereby only for fluxes
less than unity the
results of the method of self adjoint extension are reproduced
whereas for larger fluxes $N$ bound states exist and we conclude that this
method is not applicable.

We discuss the physically interesting case
of small but finite radius whereby the natural scale is given by
the anomaly of the
magnetic moment of the electron $a_e=(g-2)/2\approx 10^{-3}$.

\newpage

\section{Introduction}
There is a continuous interest in the study of scattering and
bound states  in the potential of a magnetic string
\begin{equation}\vec{A} =  {\Phi \over 2 \pi r} \vec{e}_\varphi\;,
\label{1}\end{equation}
with  the flux $\Phi$.  The interaction of a charged particle
with this potential is described by the minimal coupling
\begin{equation}\vec{p} \to \vec{p}-{e \over c} \vec{A}\;.\label{2}
\end{equation}
The corresponding magnetic field
\begin{equation}\vec{{\cal H}}=\Phi \delta (x) \delta (y) \vec{e}_3
\label{3}\end{equation}
vanishes everywhere except on the flux line where it is infinit.

The famous Aharonov-Bohm effects \cite{AB} consists in a
nontrivial scattering of a charged particle off the potential
(1). It is due to the interference of  phaseshifts  of the wave function
which are influenced by the potential (\ref{1}). In an ideal situation the
wave function vanishes where the magnetic field in nonzero demonstrating
the role of the potential. In the last years  the AB effect was studied
in connection with fractional spin and statistics in \cite{Wilczek},
its contribution to cosmic strings in \cite{Ge2}, \cite{AlfordWilczek}.
There is a close relation to the calculation of propagators in chromomagnetic
background fields \cite{Wieczorek}, \cite{Kaiser}.

While the initial investigation concerns with a scalar particle,
the inclusion of spin is natural.
In the case of a particle with spin $\vec{s}$ there is in
 addition to (\ref{2}) the interaction of its magnetic moment
\begin{equation}\vec{\mu} = g \mu_B \vec{s}/\hbar\label{4}\end{equation}
($g$ being its gyromagnetic ratio) with the magnetic field (\ref{3})
contributing
\begin{equation}\Delta \hat{H} = \vec{\mu} \vec{\cal H}\label{5}\end{equation}
to the Hamiltonian.  As it stands, this is a point interaction
and must be treated in an appropriate manner (see \cite{Albeverio,BerFadd}
for
example). From the mathematical point of view, one has to
consider the corresponding Hamilton operator on a domain of
functions vanishing on the flux line so that the term with the
$\delta$-function disappears. On this domain the operator is not
self-adjoint and its self-adjoint extensions (a one
parameter family labeled by $\lambda$) define just all possible
point interactions (\ref{3}).

In the case of a neutral particle with magnetic moment (i.e.
with the interaction (\ref{5})) this can be found in the book
\cite{Albeverio} within a general mathematical framework.
  For a spinor particle, using the Dirac
equation,  this analysis has been done in
\cite{Ge2}, \cite{Ha5,Ha1}. There it was shown that the self adjoint
extensions can be defined by
proper boundary conditions on the wave function on the flux
line. Also, an analysis using a regularized $\delta$-function
was done in \cite{Ge2} and \cite{Vo2}. Thereby the possibility of a bound
state was discussed.
Similar  results exist for the spin 1 case \cite{Ha2}.

In general, the Dirac equation
leads to a magnetic moment which is characterized by a gyromagnetic ratio
of $g=2$.
This case is exceptional from the point of view of its interaction with a
magnetic flux line because the repulsive force of the AB effect is
exactly compensated by the attractive force from the interaction of the
magnetic moment with the flux (in case when they are antiparallel).
This produces zero modes, i.e. bound states of zero binding energy.
This situation was probably first mentioned in  \cite{AC1} where it was shown
that a spin-${1\over 2}$ particle (described by a Pauli (with $g=2$) as well
as by Dirac equation)
in a (in general, nonsingular) magnetic field of total
flux ${\Phi\over (hc/e)}=N+\tilde{\delta}$,
$0<\tilde{\delta}<1$ has $N$ zero-energy normalizable eigenstates. It has the
remarkable
property, that its Hamilton operator factorizes and both equations
have essentially the same form. This is a example for a
supersymmetric quantum mechanical system.

Now, it is clear, that an anomalous magnetic moment destroyes this property.
Having in mind realistic particles like the electron with its
anomaly factor
\begin{equation} a_e\equiv{g-2\over 2}=0.001159\label{ae}\end{equation}
we consider in the present paper  a particle with spin $\vec{s}$ and
gyromagnetic ratio $g$ in 3 different, exactly solvable models of
regularization of the $\delta$-function by a flux tube of radius $R$ and
establish their connection with
the approach of self adjoint extensions.
We consider to what extend this models correspond to different extensions. In
each model
there are $N+1$ bound states in the case of
the gyromagnetic ratio $g$ being larger than two and the magnetic moment
directed anti parallel to the magnetic flux
and no bound state in the
opposite case. This is by one bound state more than there are zero modes
in the case $g=2$. When shrinking the radius of the flux tube to zero, the
gyromagnetic ratio must tend to 2 in order to have a finite bound state
energy.

The models, we use, are ($\vec{\cal H}=H(r) \vec{e_3}$)
\begin{eqnarray}
1.&H(r)={\Phi\over \pi R^2}\Theta (R-r)&\hbox{(homogeneous
magnetic field inside)}\label{6}\\
2.&H(r)={\Phi\over 2 \pi r R}\Theta (R-r)&\hbox{(magnetic field
proportional to $1/r$ inside)} \label{7}\\
3.&H(r)={\Phi\over 2 \pi R}\delta (r-R)&\hbox{(a cylindrical shell with
$\delta$-function)}\;.\label{8}
\end{eqnarray}

We consider for simplicity the nonrelativistic Hamilton operator
\begin{equation}\hat{H} = {1\over 2m}\left(\vec{p}-{e\over c}\vec{A}\right)^2
 + \vec{\mu} \vec{\cal H}\label{9}\end{equation}
with $\vec{\mu }$ given by (\ref{4}).

Due to spin conservation the magnetic interaction (\ref{5}) can be
replaced by \begin{equation}\pm g \mu _B
H(r)/2\, ,\end{equation}
$\pm$ corresponding to the spin projection to
the flux line. In the following we restrict ourselfes to the minus sign, i.e.
to the case
when the magnetic moment leads to a binding force. Further we choose $\Phi >0$;
for $\Phi <0$ the spin direction must be reversed.

The paper is organized as follows. In the next
section we consider the point interaction as self adjoint
extension and obtain boundary conditions on the wave function at
the origin. In the following section we consider magnetic flux
tubes (\ref{6})-(\ref{8}) with finite radius $R$, write down the wave
functions for
bound states and scattering states. In the fourth
section we consider the limit $R\to 0$ and establish its connection with
the self adjoint extension as well as physical
consequences. Conclusions are given in the last section.

\section{Self Adjoint Extension}
The
Schr\"odinger equation for the problem under consideration can
be written in the form
\begin{equation}\left({1\over 2m}\left(p-{e \over
c}{A}\right)^2-{ g
\mu _B H(r)\over 2}\right)\psi = E \psi  \label{Schreq}\end{equation}
where  $A$ and $H$ are given by (\ref{1}), (\ref{3})
for a infinitely thin flux tube resp. by (\ref{6})-(\ref{8}) for a finit
flux tube. After separation of the angular dependence and the
translational motion parallel to the flux tube
by
\begin{equation} \psi  (x) = \sum_{m=-\infty}^{\infty}
\psi_{m}(r)
{e^{-\dot{\imath} m \varphi}\over \sqrt{2\pi}} {e^{\dot{\imath} p_3
x_3}\over{\sqrt{2 \pi}}}\, , \end{equation}
the equation reads (for simplicity we set $p_3=0$)
\begin{equation} \left(-{1 \over r} {\partial \over
{\partial r}} r {\partial \over {\partial r}} +
{(m-\delta a(r))^2\over r^2} -{ g \delta h(r)\over 2} \right) \psi
_{m}(r) = \epsilon \;\psi _{m}(r)
\label{13}\end{equation}
with \[A= {\phi\over 2\pi r} a(r) \]
and $\delta = {\Phi \over (hc/e)}$ is the flux measured in units of the flux
quantum, $h(r)={1\over r}{\partial \over \partial r}a(r)$ is the radial
distribution of the magnetic field (it is normalized according to
$\int_0^\infty dr \,r\, h(r)=1$) and
with the energy $\epsilon ={E\over (2m/\hbar^2)}$.

Consider the case of a infinitely
thin flux tube: $a(r)=1$ and $h(r)=0\,\;(r\ne 0)$.
In the mathematical analysis one has to start with the Hamilton operator
\begin{equation} \hat{H}_0= -{1 \over r} {\partial \over {\partial r}} r
{\partial \over {\partial r}} +{(m-\delta)^2\over r^2} \label{15}\end{equation}
on a domain of functions
\begin{equation} {\cal D}(H_0)=\left\{\psi \in{\cal L}^2\left([0,\infty)
\right)\mid\psi (0)=0\right\} \label{16}\end{equation}
with the measure $dr\;r$. The eigenfunctions of this operator are
\begin{equation} \psi_{m}(r) =
J_{|m-\delta |}\left(\sqrt{\epsilon }r\right)
  \end{equation}
($m=0,\pm 1,..$ ). By means of the scalar product
\begin{equation}  \left( \varphi,H_0 \psi\right) =
\int _0^\infty dr\, r \,\varphi (r) \hat{H}_0 \psi (r) \,,\end{equation}
$\hat{H}_0$ is symmetric if
\begin{equation}  \lim_{r\to 0} \left(\varphi ^\ast r{\partial \over
\partial r} \psi - r{\partial \over \partial r}\varphi ^\ast \psi \right)
=0 \; . \end{equation}

For any $\psi \in \cal D$, there are - for the angular momentum $m=N$, where
$N$ is the integer part of the flux
\[\delta=N+\tilde{\delta}\quad (0\le\tilde{\delta}<1)\;,\]
- functions $\varphi \not\in \cal D$,
fulfilling this condition. They have the behaviour for $r\to 0$
\begin{equation} \varphi \sim r^{-\tilde{\delta }}+\lambda
r^{\tilde{\delta }}\;,
\;\quad \lambda \; \hbox{real}\; .\label{simodes}  \end{equation}
So, the domain $\cal D$ of $\hat{H}_0$ can be enlarged by these singular
modes (\ref{simodes}). The parameter $\lambda$ is arbitrary. Its dimension
is $r^{-2\delta}$.

Including this singular mode into the domain $\cal D$ of $\hat{H}_0$,
it becomes self adjoint. Thereby different choices of $\lambda$ lead to
different self adjoint extensions.

The corresponding eigenfunctions in the continuous part of the spectrum
($\epsilon=k^2$) are
\begin{equation}\psi_S(r)=J_{\tilde{\delta }}(kr)+B_N(k)\,
H_{\tilde{\delta }}^{(1)}
(kr)\,.\label{ssol}
\end{equation}
In general, the scattering amplitude is defined
by the asymptotics of the wave function
for $r\to\infty$
\begin{equation}\psi(r)\approx e^{\dot{\imath} k r \cos\varphi}+f(k,\varphi)
{e^{\dot{\imath} kr}\over \sqrt{r}}\,,
\label{as}\end{equation} where $\varphi$ is the scattering angle.
The usual Aharonov-Bohm scattering, i.e. without magnetic moment,
corresponds to the first term in rhs of (\ref{ssol})
and the scattering amplitude is well known. Expanding
\[f(k,\varphi)=\sum_{m=-\infty}^{\infty} f_m(k)
{e^{-\dot{\imath}m\varphi}\over\sqrt{2\pi}}\;,\]
its contribution reads
\begin{equation}f^{AB}_m(k,\varphi)=
{1\over\sqrt{k}}\left(e^{\dot{\imath}\pi(m-|m+\delta|)}-1\right)\,
e^{-\dot{\imath}\pi /4}\,.\end{equation}
So, the presence of the contribution of the Hankel function in rhs of
(\ref{ssol}) which describes a outgoing cylindrical wave
leads to an additional contribution to the scattering amplitude
\begin{equation}f_m(k)=f^{AB}_m(k)\,+\,\delta_{m,N}
{1\over\pi\sqrt{k}}B_N(k)\,.\label{addsc}\end{equation}
There is one eigenfunction describing a bound state with binding energy
$\kappa=-\epsilon$:
\begin{equation}\psi_B(r)=K_{\tilde{\delta}}\left(\sqrt{\kappa}r\right)\,.\label{bsol}
\end{equation}

Now, from the expansion of the solutions (\ref{bsol}) and (\ref {ssol}) for
$r\to 0$ we obtain by means of (\ref{simodes})
the connection of the bound state
energy $\kappa$ with the parameter $\lambda$ of the self adjoint extension:
\begin{equation}\lambda=-{\Gamma (1-\tilde{\delta})\over \Gamma
(1+\tilde{\delta})}
\left(\sqrt{\kappa}/2\right)^{2\delta}\,.\label{con1}\end{equation}
{}From this formula it follows that the bound state occurs
in the case of negative
parameter of the extension $\lambda$ only. For the scattering states
we obtain from (\ref{as}) and (\ref{simodes})
\begin{equation}B_N(k)={\dot{\imath}  \sin\pi\tilde{\delta}\over
e^{-\dot{\imath}\pi\delta}+{\Gamma (1+\tilde{\delta})\over\Gamma
(1-\tilde{\delta})}\,
\lambda (k/2)^{2\tilde{\delta}}}\,. \label{con2}\end{equation}
So, for any parameter of the parameter $\lambda$ of the extension, there
is an additional scattering and for $\lambda<0$ there is a bound state.
In that latter case the scattering amplitude can be expressed in terms
of the bound state energy
\begin{equation}B_N(k)={\dot{\imath} \sin\pi\tilde{\delta}\over
e^{-\dot{\imath}\pi\tilde{\delta}}-
\left({\kappa\over k^2}\right)^{\tilde{\delta}}}\,.\label{27}\end{equation}

\section{Three Models}

The regularization of the $\delta$-function interaction can be done by many
different
mo\-dels for a finite flux tube. We consider here the simplest examples, which
 are
exactly solvable. We write down the wavefunction inside the tube and stick
them to the outside function.

The outside function ($r>R$) is a eigenfunction of the Hamilton
operator (\ref{15}).
 For $\epsilon<0$ it is given by
\begin{equation}\psi_m(r)=K_{m-\delta}(\sqrt{-\epsilon}r)\,\label{27.1}
\end{equation}
and describes the bound state solution. Its logarithmic derivative reads
\begin{equation}R_{ex}\equiv R{\partial\over\partial r}
\ln{\psi_m(r)}_{\mid _{r=R+0}}=
\sqrt{-\epsilon}R{K'_{m-\delta}\left(\sqrt{-\epsilon}R\right)\over
K_{m-\delta}\left(\sqrt{-\epsilon}R\right)}\,.\label{28}\end{equation}
For $\epsilon=k^2>0$ we obtain the outside scattering solution ($r>R$)
\begin{equation}\psi_m(r)=J_{|m-\delta|}(kr)+B_m(k)\,H_{|m-\delta|}^{(1)}(kr)\label{28.1}\end{equation}
and its logarithmic derivative reads
\begin{equation}R_{ex}\equiv r{\partial\over\partial r}
\ln{\psi_m(r)}_{\mid _{r=R+0}}=
\sqrt{\epsilon}R {
    J'_{|m-\delta|}\left(\sqrt{\epsilon}R\right)+
    B_m(k) \, H^{(1)'}_{|m-\delta|} \left( \sqrt{\epsilon}R \right)
\over
J_{|m-\delta|} \left(  \sqrt{\epsilon}R \right)+
B_m(k) \, H^{(1)}_{|m-\delta|} \left( \sqrt{\epsilon}R \right)
} \,.  \label{29}\end{equation}
Below we are interested in the limit $R\to 0$. For $\epsilon<0$ we note
\begin{eqnarray}
R_{ex}&=&\left\{ {
-|m-\delta|-2|m-\delta|{\Gamma (1-|m-\delta|)
\over\Gamma (1+|m-\delta|)}
\left({\sqrt{-\epsilon}R\over 2}\right)^{2|m-\delta|}+\,...
\quad |m-\delta|<1
\atop
-|m-\delta |-2{1\over |m-\delta |-1}
\left({\sqrt{-\epsilon}R\over 2}\right)^{2}+\,...
\quad|m-\delta|>1\,,   }\right.
\label{a}\end{eqnarray}
where two cases have to be distinguished.

\subsection{Homogeneous Magnetic Field}

In this model the magnetic field inside is homogeneous and zero outside. The
functions
$h(r)$ and $a(r)$ read:
\begin{equation}h(r)={2\over R^2} \Theta (R-r),\quad a(r)={r^2\over R^2}
\Theta (R-r)\,+\,\Theta (r-R)\,.\end{equation}
The Schr\"odinger equation reads
\begin{equation}\left(
-{1\over r}{\partial\over\partial r}r{\partial\over\partial r} +
   {   \left( m-\delta {r^2\over R^2}  \right)^2 \over r^2}
   - {g\over 2}\delta {2\over R^2}
\right)  \psi_m(r)=\epsilon \psi_m(r)\,.\end{equation}
The solution regular in $r=0$ is given by
\begin{equation}\psi_m(r)=r^{|m|}\,
_1F_1\left(  {2-g\over 4}+{|m|-m\over 2}- {\epsilon R^2\over 2\delta },1+|m|,
{\delta r^2\over R^2}  \right) e^{-{\delta r^2\over 2R^2}}\,.
\label{33}\end{equation}
We need  its logarithmic derivative at $r=R$
\[R_1\equiv R{\partial\over\partial r}{\psi_m(r)}_{\mid _{r=R-0}}\]
for $m\ge 0$ and use the notation $x\equiv \sqrt{-\epsilon}R$:
\begin{equation}R_1=\,|m|-\delta+\delta {{2-g\over 4}+{x^2\over 4\delta}
\over 1+|m|}\,
{_1F_1({2-g\over 4}+1+{x^2 \over 4\delta},2+|m|;\delta)\over
_1F_1\left({2-g\over 4}+{x^2\over 4\delta},1+|m|;
\delta\right)}\,.\end{equation}
For $x\to 0$ we note
\begin{equation}R_1=|m|-\delta+{2-g\over 2}\delta m\alpha_1+x^2
\beta_1+ \,...\label{b}\end{equation}
with \[\alpha_1={1\over 2(1+|m|)} {_1F_1({2-g\over 4}+1,2+|m|;
\delta)\over
_1F_1({2-g\over 4},1+|m|;\delta)}\;,
\beta_1={\partial\over\partial g}{2-g\over 2}\alpha_1\;.\]
The properties $\alpha_1>0$ and $\beta_1<0$ can be checked.

\subsection{Magnetic Field Proportional to $1/r$}

In this model we have
\begin{equation}h(r)={1\over rR} \Theta (R-r),\quad a(r)={r\over R}
\Theta (R-r)\,+\,\Theta (r-R)\,.\end{equation}
The corresponding equation reads
\begin{equation}\left(
-{1\over r}{\partial\over\partial r}r{\partial\over\partial r} +
   {   \left( m-\delta {r\over R}  \right)^2 \over r^2}
   - {g\over 2}\delta {1\over rR}
\right)  \psi_m(r)=\epsilon \psi_m(r)\,.\end{equation}
and it has a solution regular in $r=0$
\begin{equation}\psi_m(r)=r^{|m|}\, _1F_1\left(
m\left({1\over 2}+|m|-m{\delta\over\bar{\delta}}\right)
-{g\over 4}{\delta\over\bar{\delta}},1+2|m|;2\bar{\delta}
{r\over R}
\right)\, e^{-\bar{\delta}{r\over R}}
\,\label{37.1}\end{equation}
with the notation $\bar{\delta}\equiv\sqrt{\delta^2-\epsilon R^2}$.
Its logarithmic derivative reads
\[R_2\equiv r{\partial\over\partial r}{\psi_m(r)}_{\mid _{r=R-0}}\]
\[=\,|m|-\bar{\delta}+2{ ({1\over 2}+|m|)\bar{\delta}-(m+g/4)\delta
 \over 1+2|m|}
\,{ _1F_1\left({3\over 2}+|m|-(m+g/4){\delta \over\bar{\delta}}
,2+2|m|;2\bar{\delta}\right)\over
_1F_1\left({1\over 2}+|m|-(m+g/4){\delta
\over\bar{\delta}},1+2m;2\bar{\delta}\right)
}\,.\]
For $m\ge 0$ and $x\equiv \sqrt{-\epsilon }R\to 0$ we note
\begin{equation}R_2=m-\delta+{2- g\over 2}\delta \alpha_2+\beta_2\,x^2+\,
...\label{b2}\end{equation}
with
\[\alpha_2={1\over 1+2m}{_1F_1\left({2-g\over 4}+1,2+2m;2\delta\right)\over
_1F_1\left({2-g\over 4},1+m;2\delta\right)}\]
and \[\beta_2={1\over 2\delta}\left((1+2m)\alpha_2-1+
{2-g\over 2}\left({\partial\over\partial\delta}-
(g+4m){\partial\over\partial g}\right)\alpha_2\right)\,.\]
Also in this case the properties $\alpha_2>0$ and $\beta_2<0$ can be checked.

\subsection{Cylindrical Shell with $\delta$-function}
Moving the $\delta$-function from $r=0$ to $r=R$ one obtains a cylindrical
shell
on which the magnetic field is infinite\footnote{This model is intensively used
in \cite{Ha5}.}:
\[a(r)=\Theta(R-r),\quad h(r)={1\over R}\delta (r-R)\,.\]
The radial equation reads
\begin{equation}\left(
-{1\over r}{\partial\over\partial r}r{\partial\over\partial r} +
   {   \left( m-\delta \;\Theta (R-r)  \right)^2 \over r^2}
   - {g\over 2}\delta {1\over R}\delta (r-R)
\right)  \psi_m(r)=\epsilon \psi_m(r)\,.\label{eqcyl}\end{equation}
In this case the $\delta$-function because moved away from the origin can be
treated as usual in 1-dimensional case
and substituted by the known boundary conditions
\begin{equation}r \partial_r \psi_m (r) \big|^{^{R+0}}_{_{R-0}}=-
{g\over 2}\delta\psi_m
(r)_{|_R}\,. \label{bbed3}\end{equation}
Then the solution of eq. (\ref{eqcyl}) are Bessel functions
\begin{equation}\psi_m(r)=\left\{\alpha J_{|m|}(\sqrt{\epsilon}r)\quad
\hbox{for}\;r<R
\atop J_{|m-\delta|}(kr)+B_m(k)\,H_{|m-\delta|}^{(1)}(kr) \quad \hbox{for}\;r>R
\right.\label{39.1}\end{equation}
with some coefficient $\alpha$ and from condition (\ref{bbed3}) it follows
\[\begin{array}{lll}
R_3&\equiv
- {1\over 2}\,g\,\delta \,+\,r{\partial\over\partial r}
{\psi_m(r)}_{\mid _{r=R-0}}& \\[5pt]
&=\,- {1\over 2}\,g\,\delta \,+\,\sqrt{\epsilon}R{J_{|m|}'\left(\sqrt{\epsilon}
R\right)\over
J_{|m|}\left(\sqrt{\epsilon}R\right)}\quad &\hbox{for} \quad\epsilon>0\\[8pt]
&=\,- {1\over 2}\,g\,\delta \,+\,\sqrt{-\epsilon}R{I_{|m|}'\left(\sqrt{-
\epsilon}R\right)\over
I_{|m|}\left(\sqrt{-\epsilon}R\right)}\quad &\hbox{for}\quad \epsilon<0\,.
\end{array}\]
For $x\equiv\sqrt{-\epsilon} R\to 0$ we have
\begin{equation}R_3=|m|-\delta+{2- g\over 2}\delta \alpha_3+
\beta_3 x^2 +\,...\label{b3}\end{equation}
with
\[\alpha_3=1\,,\;\beta_3={1\over 2(1+|m|)}\,.\]

\subsection{Bound State Energy and Scattering Amplitude}

The solutions in all three models are determined by the condition
\begin{equation}R_{ex}=R_i \quad (i=1,2,3)\,.\label{bed1}\end{equation}
There are scattering solutions for all values of the parameters. They
can be obtained by solving (\ref{bed1}). The scattering amplitude reads
\begin{equation}B_m(k)=-{\left( x{\partial\over\partial x}-R_i\right)
J_{|m-\delta|}(x)
\over \left( x{\partial\over\partial x}-R_i\right) H^{(1)}_{|m-\delta|}(x)
}_{|x=kR}\,.\label{scam}\end{equation}

 Let us consider the
bound state  solutions.
They does not exist for all values of the parameters. Consider the behaviour
of $R_i$ and $R_{ex}$ as function of $x\equiv\sqrt{-\epsilon}R$. It can be
seen that $R_{ex}$ decrease starting from $R_{ex}(0)=-|m-\delta|$ (cf.
(\ref{a})) while $R_i(x)$ increase starting from
$R_i(x)=|m|-\delta+{2-g\over 2}\delta m\alpha_i$ (cf. (\ref{b}),
(\ref{b2}), (\ref{b3}).
So, solutions with binding energy $\kappa_m\equiv\sqrt{-\epsilon}=x/R$
of eq. (\ref{bed1}) are possible for
\begin{equation}g>2,\quad 0\le m<\delta\left(1+{g-2\over 4}
\alpha_i\right)\,.\label{bed2}\end{equation}
In the case $g=2$ all solutions have vanishing energy, i.e. they correspond
to zero modes.
In general, the solution of (\ref{bed1}) reads
\begin{equation}x=f(\delta,g,m)\quad \hbox{with}\quad x
=\kappa_mR\,,\label{41.1}\end{equation}
where $f$ is some dimensionless function. Some lowest solutions of
eq. (\ref{bed1}) are shown in the figure for the model with the cylindrical
$\delta$-shell.
\unitlength1cm
\begin{picture}(14,10)
\put(0.5,1){\parbox{14cm}{\small {\bf Fig. } \quad The solutions $x=f(d,g,m)$
 of eq. (\ref{41.1}) for $m=0,1,\,...\,4$ and $g=2.2,\,2.1,\,2.05,\,2.01$
}}
%\put(0.5,-2){\psfig{figure=abb.ps,height=15cm,width=14cm}}
\end{picture}\\
Similar pictures can be drawn for the other two models.
It can be seen that, in general,  there is no simple rule
for the energy levels $\kappa_m=x/R$ for general values of the parameters.

\section{The Limit $R\to 0$}

In the limit $R\to 0$, all other parameters fixed, the bound state energy
increases unbounded as can be seen from (\ref{41.1}).
This indicates, that the limit $R\to 0$ in the models
with finite flux tube is not physical at last in the nonrelativistic
approximation choosen here.

One possibility to obtain a finite energy is to consider a small anomaly
of the magnetic moment, i.e. to consider the case
\[a_e\equiv {g-2\over 2}\to 0\,.\]
In that case all solutions $x$ of eq. (\ref{41.1}) tend to zero and
 the equation (\ref{bed1}), which defines the bound states, can be solved.
 We obtain, using (\ref{a}) and (\ref{b}), (\ref{b2}), (\ref{b3}),
for
the highest angular momentum
\begin{equation}{g-2\over 2}\delta\alpha_i=\left({\sqrt{\kappa_N}R\over 2}
\right)^{2
\tilde{\delta}} {2\tilde{\delta} \Gamma (1-\tilde{\delta})\over \Gamma (1+
\bar{\delta})}\quad (m=N),\label{en1}\end{equation}
where $N$ is the integer part of the flux,
and
\begin{equation}{g-2\over 2}\delta\alpha_i=(\sqrt{\kappa_m}R)^2\left(
\beta_i+{1\over
4(|\delta-m|-1)}\right)\quad (m=0,1,\,...\,,N-1) \,\label{en2}\end{equation}
for the lower angular momenta.

For $g\to 2$, $R$ fixed, $\kappa_m$ ($m=0,1,\,...\,N-1$) tends to zero
proportional to $g-2$, whereas $\kappa_N$ behaves like
$(g-2)^{1/\bar{\delta}}$, i.e. tends more quickly to zero.
Therefore, $\kappa_m$ correspond to the zero modes (for $g=2$), whereas the
state with $\kappa_N$ has no correspondence in that case.
It can be expected that its wave function vanishes.

For $R\to 0$,  in the case $N=0$, i.e. the flux beeing less than unity,
a finite binding energy of the only bound state can be obtained by
substituting
\begin{equation}{g-2\over 2}\to{2R^{2\delta}\over\alpha_i}\,{g-2\over  2}
\label{sub}\end{equation}
in the initial equation (\ref{Schreq}) and after that performing the limit
$R\to 0$.
After the substitution (\ref{sub}), the bound state energy is determined by
(instead of (\ref{en1}))
\begin{equation}\left({\sqrt{\kappa_0}\over 2}\right)^{2\delta}
=\,{\Gamma (1+\delta)\over\Gamma (1-\delta)}\,{g-2\over 2}\label{sub1}
\end{equation}
and we observe from eq. (\ref{con1}) that the parameter $\lambda$ of the
extension is just (up to the sign) the anomaly of the magnetic moment:
\begin{equation}\lambda=-{g-2\over 2}\,.\label{sub2}\end{equation}
This treatment of the $\delta$-function is equivalent to the general
approach to 2-dimensional $\delta$-function in the Schr\"odinger equation
 by Berezin, Faddeev \cite{BerFadd} and Albeverio \cite{Albeverio},
where the necessarity of the renormalization of the coupling was pointed out.

In the case of fluxes larger than unity ($N\ge 1$), there are bound states
with energy $\kappa_m$ larger than $\kappa_N$
and the renormalization (\ref{sub}) is not sufficient to keep them finite.
Instead, by means of (\ref{en2}), one must substitute
\begin{equation}{g-2\over 2}\to{g-2\over 2}R^2\label{sub3}\end{equation}
in the initial equation (\ref{Schreq}).
Than the binding energies read (instead of (\ref{en2}))
\begin{equation}\kappa_m={g-2\over 2}{\delta\alpha_i\over
\beta_i+1/(4(|\delta -m|-1))}\label{sub4}\end{equation}
($m=0,1,\,...\,N-1$). In this case we have $\kappa_N=0$.

A different way of understanding the limit $R\to 0$ is to keep $R$
small, but finite.  In that case there is a natural scale given at the one
hand side by the value of the anomaly of the magnetic moment of the electron
\[a_e={g-2\over 2}=0.001159 \]
and on the other hand side by the bound state energy to be nonrelativistic,
i.e. smaller than the electron mass because we consider the nonrelativistic
Schr\"odinger equation.

In the case of flux less than unity the energy is nonrelativistic for
$\kappa_0<<1/\lambda_c$, the inverse Compton wavelength of the electron,
and the radius must  by means of eq. (\ref{en1}) fulfill
\begin{equation}R>>\left({g-2\over 2}{\Gamma (1+\delta )\over 2\Gamma (1-\delta
)}\alpha_i\right)^{1/(2\delta)} 2\lambda_c\label{gr1}\end{equation}
Thereby it can be taken smaller than $\lambda_c$ so that the flux tube can be
considered as thin. Lets remark, that the considerations  done here for small
$x=\sqrt{\kappa} R$ mean that the size of the orbit
of the bound states is much larger than the radius $R$.

Similar considerations apply to the case of the flux  being larger than unity.
Here, the radius must fulfill
\begin{equation}R>>\left({g-2\over
2}\delta{1\over\beta_i+1/4(|\delta-m|-1)}\alpha_i\right)^{1/2}\lambda_c\label{gr2}\end{equation}
This condition is  stronger  than (\ref{gr1}). Nevertheless, $R$ may be made
smaller
than $\lambda_c$, so that the flux tube can be thin in this case too.

Let's consider the limit $R\to 0$ for  scattering states, i.e.
$k^2=\epsilon>0$.
Expanding $B_m(k)$ (\ref{scam}), i.e. the additional scattering amplitude
for a given angular momentum $m$ and energy $k^2$, for $x\to 0$:
\begin{equation}B_m(k)={\dot{\imath}\sin\pi\nu\,\left(
{x\over 2}\right)^{2\nu} \over
{\Gamma (1+\nu)\over \Gamma (1-\nu)(2\nu+{g-2\over 2}\delta\alpha_i)}
\left(1+\left(\beta_i-{{2-g\over 2}\delta\alpha_i-2\over 8(1-\nu)}\right)
x^2\right)+\left({x\over 2}\right)^{2\nu}
e^{-\dot{\imath}\pi\nu}
}\label{b1}\end{equation}
with $\nu=|m-\delta|$. From this formula it can be seen, that $B_m(k)$
vanishes in the limit $R\to 0$, all other parameters fixed. This is
meaningful in the case $g<2$ where there are no boundstates.

For $g>2$, as shown above, in order to have finite bound state energies,
the limit $R\to 0$
must be performed together with $g\to 2$.

For $0\le\delta<1$, one must use the substitution (\ref{sub}) and obtains
\begin{equation}B_0(k)\approx {\dot{\imath}\sin\pi\nu
\over
e^{-\dot{\imath}\pi\nu}-{g-2\over 2}\delta{\Gamma (1+\nu)\over \Gamma
(1-\nu)}\left({k\over 2}\right)^{-2\delta}}\,,\label{B0}\end{equation}
i.e. the same formula as in the method of self adjoint extensions (\ref{con2}).

For flux larger than one, i.e. $\delta=N+\bar{\delta}>1$, one has to use
the substitution (\ref{sub3}) and obtaines
\begin{equation}B_N(k)={1\over 2}\left(e^{2\dot{\imath}\pi\bar{\delta}}
-1\right)\quad (m=N)\label{62}\end{equation}
and
\begin{equation}B_m(k)\approx R^{2(\delta -m-1)}\to 0 \quad (m=0,1,\,...
\,N-1)\label{63}\end{equation}
So, the additional scattering takes place for the highest angular momentum
only.

\section{Conclusions}

We considered a charged quantum mechanical particle with spin
${1\over 2}$ and gyromagnetic ratio $g\ne 2$ in the field af a
magnetic string. Whereas the interaction of the charge with the
string is the well kown Aharonov-Bohm effect and the
contribution of the magnetic moment associated with the spin in the
case $g=2$ is known to yield an additional scattering and zero
modes (one for each flux quantum), an anomaly of the magnetic
moment (i.e. $g>2$) leads to bound states. We considered two
methods for treating the case $g>2$.

For a ideal string the interaction of the spin with the magnetic
field (\ref{5}) is pointlike and singular; i.e. the magnetic
field containes the 2-dimensional $\delta$-function (\ref{3}).
A mathematical approach to treat this is the method of self
adjoint extension. It yields a family of operators labeled by a
real parameter $\lambda$. For all values of this operator there
is an additional scattering amplitude (\ref{addsc}) resulting
from the contribution of the magnetic moment and for $\lambda<
0$ there is one bound state (\ref{con1}). The main goal of the
extension is to include a singular solution (\ref{ssol}) resp.
(\ref{bsol}) into the domain of the Hamilton operator
(\ref{15}). It should be remarked that this method is - although
mathematically correct (or, at last, may be made correct) - not
satisfactory from the physical point of view because the
parameter $\lambda$ of the extension is not correlated to physical
parameters like the gyromagnetic ratio which does not enter this
method at all.

A different method is to consider nonsingular flux tubes and
shrinking its radius to zero. This is equivalent to regularize
the $\delta$-function in the magnetic field by some less
singular profile. We used 3 models for which the Schr\"odinger
equation (\ref{Schreq}) (it is really a Pauli equation in this
case) can be solved explicitely. The common result is that there
is an additional scattering due to the magnetic moment and that
for $g>2$ there are bound states. This is not surprisingly since
the existence of zero mode for $g=2$ suggests the binding of a
particle as soon as there is any additional attrative force. For
$g$ slightly larger than $2$ there are $N+1$ bound states, where
$N$ is the integer part of the flux. In general, the dependence
of the energy levels $\kappa_m$ on $g$ and on the flux is complicated as
can be seen from the figure.

We considered the case $g\to 2$. Thereby all $\kappa_m$ tend to
zero. For $m=0,1,\,...\,N-1$, the solutions become the mentioned
zero modes, for $m=N$ the solution vanishes.

The limit $R\to 0$, all other parameters fixed, is not physical
for $g>2$. The reason is that the bound state energies
$\kappa_m$ enter the defining equation (\ref{41.1}) multiplied
by the flux tube radius $R$, which follows from general
dimensional considerations too. So, $\kappa_m$ tend to infinity
for $R\to 0$.

One possibility to treat this problem is to tend the
gyromagnetic ratio $g$ in the initial equation to $2$ along with
$R\to 0$. This can be viewed as some renormalization of $g$. Thereby two cases
have to be distinguished. Firstly,
when the flux is less than unity, $g-2$ tend to zero
proportional to $R^{2\delta}$, eq. (\ref{sub}). For $g>2$, there is one bound
state, its energy is given by eq. (\ref{sub1}). This is essentially
the same situation as in the method of self adjoint extension
and the parameter $\lambda$ of the extension can be related to
the gyromagnetic ratio, eq. (\ref{sub2}). Thereby the dependence
on the parameters of the models used enters the renormalization
(\ref{sub}) of the gyromagnetic ratio only. Such a renormalization
is known in the mathematical approach \cite{BerFadd}, too. For
the scattering amplitude there is a contribution in addition to
the usual Aharonov-Bohm scattering (eq. (\ref{B0})). It is given
by the same formula as in the method of extension, eq.
(\ref{27}). So, for $\delta <1$ both approaches are equivalent.

A different feature one observes for flux larger than unity. In
order to keep all bound state energies finite, one is forced to
tend $g-2$ to zero proportional to $R^2$, eq. (\ref{sub3}),
i.e. much faster than in the previous case. Thereby the energy
of the state $\kappa_N$ (with the highest angular momentum
$m=N$) tends to zero whereas for $m=0,1,\,...\,,N-1$ the energies
$\kappa_m$ are finite. The additional scattering in this case
takes place for $m=N$ only, eq. (\ref{62}); for
$m=0,1,\,...\,,N-1$, $B_m(k)$ (eq. (\ref{63})) vanishes for fixed parameters.
This is clear because the wave functions in this case are concentrated
in the region of small $R$. For sufficiently high momenta $k$ scattering
can be expected in this case too.
So we conclude, that for flux larger than one both approaches yield
different results. This must be not a contradiction by the following reason.
In the method of self adjoint extension, the input information, which is
contained in the Hamilton operator (\ref{15}) makes no reference to the
magnetic moment of the particle as well as not to the magnetic field and,
moreover, no reference to the integer part $N$ of the magnetic flux by
means that it enters in the combination $m-\delta$ only. So, as from the
other method it is known, that the existence of boundstates requires $g>2$.
In the same manner we conclude that the extension is applicable for flux
less than one only, i.e. that a flux larger than one is too singular and
cannot be described by the method of self adjoint extension.

The physically interesting case is to keep the flux tube radius
small, but finite. The natural scale for the smallness of $g-2$
is given by the anomaly $a_e=(g-2)/2$ (\ref{ae}) of the magnetic
moment of the electron and by the requirement the radius of the
tube being not too small in order to have nonrelativistic bound
state energies $\kappa_m<<m_e$ ($m_e$ - the electron mass)
because we use a nonrelativistic equation. Under this
conditions there are approximately $N+1$ bound states. The
exact number is given by eq. (\ref{bed2}) and depends on the
model used. However, this dependence is weak for small $a_e$ as
can be seen from eq. (\ref{bed2}) where $\alpha_i$ enters
multiplied by $g-2$. Furthermore, it should be emphasized that
the physical restrictions to $R$ and $g-2$ allow the flux tube
to be thin in that sense that $\kappa_mR<<1$ is possible, i.e.
the orbit size of the bound states is much larger than the
radius of the flux tube.

So, we conclude that for a gyromagnetic ratio larger than $2$
the flux tube cannot be shrinked to a line for real physical
parameters. A natural extension of these investigations would be
the consideration of the Dirac equation with additional magnetic
moment (i.e. including a term $(g-2)\sigma^{\mu\nu}F_{\mu\nu}$).
In that case, the limitation to the bound state energy being
nonrelativistic is not necessary and smaller $R$ can be
considered. Furthermore, one can speculate that the anomaly of
the magnetic moment, which is known to decrease in strong magnetic fields
\cite{Gu},  will eventually influnece the limit $R\to 0$.

A further open question is, wheather the interaction, which comes
from the anomaly $a_e$ of the magnetic moment, can be treated in
perturbation theory with respect to $a_e$ starting from the
known solutions (especially from the zero mode of \cite{AC1}) for
an arbitrary profile of the magnetic field inside a finite flux
tube. \\[1cm]
{\small The authors thank J. Audretsch, E. Seiler, and E. Wieczorek
for several discussions and helpful suggestions. One of us ( S. V. ) would
like to thank the NTZ of Leipzig University for kind hospitality.}

\end{document}